\documentclass[journal, letterpaper]{IEEEtran}

\usepackage{graphicx}

\usepackage{algorithm}
\usepackage{algpseudocode}
\usepackage{tikz}
\usepackage{pgfplots}
\pgfplotsset{compat=1.17}

\usepackage{url}

\usepackage{amsmath,amsfonts,amssymb,amsthm}

\usepackage{textgreek}	% Greek to me, dawg
\usepackage{listings}
\usepackage{csvsimple}
\usepackage{longtable}

\begin{document}

	\title{Quantum Reinforcement Learning in Non-Abelian
 Environments: Unveiling Novel Formulations and Quantum Advantage Exploration}
	\author{Shubhayan Ghosal \\  Department of Mathematics, Jadavpur University \\  Email: \texttt{shubhayang.math.ug@jadavpuruniversity.in}}
	\markboth{}{k}
	\maketitle

\begin{abstract}

This paper delves into recent advancements in Quantum Reinforcement Learning (QRL), particularly focusing on non-commutative environments, which represent uncharted territory in this field. Our research endeavors to redefine the boundaries of decision-making by introducing  formulations and strategies that harness the inherent properties of quantum systems.

At the core of our investigation characterization of the agent's state space within a Hilbert space ($\mathcal{H}$). Here, quantum states emerge as complex superpositions of classical state introducing non-commutative quantum actions governed by unitary operators, necessitating a reimagining of state transitions. Complementing this framework is a refined reward function, rooted in quantum mechanics as a Hermitian operator on $\mathcal{H}$. This reward function serves as the foundation for the agent's decision-making process. By leveraging the quantum Bellman equation, we establish a methodology for maximizing expected cumulative reward over an infinite horizon, considering the entangled dynamics of quantum systems.We also connect the Quantum Bellman Equation to the Degree of Non Commutativity of the Environment, evident in Pure Algebra.

We design a quantum advantage function. This ingeniously designed function exploits latent quantum parallelism inherent in the system, enhancing the agent's decision-making capabilities and paving the way for exploration of quantum advantage in uncharted territories. Furthermore, we address the significant challenge of quantum exploration directly, recognizing the limitations of traditional strategies in this complex environment.

\end{abstract}
\section{Introduction}
	
In quantum mechanics, the state of a quantum system is described by a vector in a complex vector space called a Hilbert space, denoted as $\mathcal{H}$. This space contains all possible states of the system. Each vector in $\mathcal{H}$ represents a possible state of the system, and these states can be superpositions of classical states, meaning they can exhibit both wave-like and particle-like behavior simultaneously.

Now, when we consider an agent operating in a quantum environment, its actions are described by unitary operators acting on the quantum state. A unitary operator is a linear transformation that preserves the inner product and norm of vectors, meaning it preserves the probabilities of outcomes upon measurement. Mathematically, for a given action $a$, the agent applies a unitary operator $U_a$ to its current state $\psi_t$ to transition to a new state $\psi_{t+1}$.

However, an important property of quantum actions is that they are non-commutative. This means that the order in which the actions are performed matters. In general, for two unitary operators $U_a$ and $U_b$, $U_a U_b$ is not equal to $U_b U_a$. This non-commutativity is a fundamental aspect of quantum mechanics and is a consequence of the uncertainty principle.

To elaborate, consider two quantum actions represented by unitary operators $U_a$ and $U_b$. Performing action $a$ followed by action $b$ is not the same as performing action $b$ followed by action $a$. In other words, the effect on the quantum state depends on the order in which the actions are applied. This non-commutativity can lead to interesting and sometimes counterintuitive behavior in quantum systems, and it plays a crucial role in quantum computation, quantum information theory, and quantum mechanics in general.\cite{ryan2019}

In classical mechanics, the order in which operations are performed typically does not matter. For example, if you push a box to the left and then push it forward, or if you push it forward and then to the left, the final position of the box will be the same. This property is known as commutativity.

However, in quantum mechanics, the situation is fundamentally different due to the principles of superposition and uncertainty. Quantum systems can exist in superpositions of states, meaning they can be in multiple states simultaneously. When we apply operations to these systems, the non-commutativity of quantum actions arises due to the nature of quantum operators and the uncertainty principle.

\subsection*{Definition:}
A module generalizes the concept of vector spaces and is equipped with scalar multiplication from a ring. In an abelian module, scalar multiplication commutes, but in a non-abelian module, it does not necessarily commute.

\subsection*{Non-Abelian Modules in Quantum Context:}
In quantum computing, the actions of quantum operators (e.g., quantum gates) can form non-abelian modules. This arises notably in the context of topological quantum computation.

\subsection*{Role in Quantum Information Processing:}
\begin{itemize}
    \item \textbf{Non-Abelian Statistics}: Certain quasiparticles, such as anyons in topological states, exhibit non-abelian braiding statistics. Quantum information can be encoded in states with multiple quasiparticles, leveraging non-abelian braiding for computation.
    \item \textbf{Topological Quantum Computation}: Fault-tolerant quantum computation can be achieved using non-abelian anyons and their braiding operations. Topological quantum computing harnesses the topological properties of these systems to protect quantum information from errors.

    \item \textbf{Braiding Operations}: Quantum algorithms can exploit the non-abelian properties of modules, particularly in the context of topological quantum computation. Braiding operations on non-abelian anyons encode and process quantum information, offering potential advantages over classical computation.
    \item \textbf{Fault Tolerance}: The fault-tolerance of topological quantum computers relies on non-abelian properties to encode and protect quantum information from errors.
\end{itemize}

Let's consider two unitary operators, \(U_a\) and \(U_b\), corresponding to two quantum actions. When we apply \(U_a\) followed by \(U_b\) to a quantum state \(\psi\), we get \(U_b(U_a(\psi))\). Similarly, when we apply \(U_b\) followed by \(U_a\), we get \(U_a(U_b(\psi))\). This non-commutativity leads to interesting consequences:
\begin{enumerate}

\item \textbf{Measurement}: When the agent performs a measurement on its quantum state $\psi$, it collapses to one of the eigenstates of the measured observable with probability determined by the Born rule.
In quantum mechanics, the act of measuring a quantum system is a fundamental concept with profound implications. When a measurement is performed on a quantum system that is in a superposition of states, it causes the system to transition from a state of potentiality to a definite state. This transition is known as the collapse of the wavefunction.

\begin{enumerate}
    \item \textbf{Quantum State}: A quantum system is described by a mathematical object called a wavefunction or state vector. The state vector represents all the possible states the system can be in. Mathematically, this state vector can be represented by the symbol $\psi$.
    
    \item \textbf{Superposition}: Prior to measurement, a quantum system can exist in a superposition of multiple states. For instance, an electron can be in a superposition of spin-up and spin-down states.
    
    \item \textbf{Measurement Process}: When a measurement is performed on the quantum system, it interacts with a measuring device. This interaction leads to the emergence of classical behavior. However, the act of measurement itself is inherently probabilistic.
    
    \item \textbf{Eigenstates and Eigenvalues}: In quantum mechanics, observables such as position, momentum, energy, and spin are represented by Hermitian operators. When a measurement is made on an observable, the state of the system collapses to one of the eigenstates of that observable. Each eigenstate corresponds to a possible outcome of the measurement, and the corresponding eigenvalue represents the value that will be measured.
    
    \item \textbf{The Born Rule}: The Born rule, formulated by physicist Max Born, provides a probabilistic interpretation of the measurement outcome. According to the Born rule, the probability of obtaining a particular measurement outcome is proportional to the square of the absolute value of the probability amplitude associated with the corresponding eigenstate.
    
\end{enumerate}

Mathematically, if $\psi$ is the state of the quantum system, and $\hat{A}$ is the operator representing the observable being measured, with eigenstates $|a_i\rangle$ and corresponding eigenvalues $a_i$, then the probability $P(a_i)$ of obtaining the measurement outcome $a_i$ when measuring the observable $A$ is given by:

\[
P(a_i) = |\langle a_i | \psi \rangle|^2
\]

After the measurement, the quantum system is left in the eigenstate corresponding to the observed eigenvalue. This is known as the collapse of the wavefunction.

It's important to note that the collapse of the wavefunction is a conceptually challenging aspect of quantum mechanics, and various interpretations exist to explain it, including the Copenhagen interpretation, many-worlds interpretation, and the objective collapse theories, among others.

\item \textbf{Entanglement}: Quantum reinforcement learning models may exploit entanglement between the agent's state and other quantum systems, potentially leading to novel learning dynamics and capabilities, described using the concept of a joint quantum state that cannot be factored into individual states for each particle.In quantum reinforcement learning (QRL) models, the exploitation of entanglement involves representing the agent's state and other relevant quantum systems in an entangled manner. Mathematically, we can describe this entanglement by considering a joint quantum state involving both the agent's state and additional quantum systems\cite{chen2020}

Let's denote the quantum state of the agent as \(|\psi_A\rangle\), and the quantum state of the environment or other relevant systems as \(|\psi_E\rangle\). In entangled QRL models, the joint quantum state of the agent and environment is represented as an entangled state, denoted by \(|\Psi\rangle\), which cannot be factorized into separate states for the agent and environment:

\[|\Psi\rangle = |\psi_A\rangle \otimes |\psi_E\rangle\]

However, this representation may oversimplify the entanglement present in QRL. In more complex scenarios, entanglement may arise between various components of the QRL system, including the agent's state, the environment, and possibly auxiliary quantum systems used for computation or measurement.This could involve representing the QRL system as a multipartite quantum state involving multiple qubits or quantum registers. The entanglement between different parts of the system would then be characterized by the correlations present in this multipartite quantum state.

Quantum entanglement in QRL models introduces non-trivial correlations between the agent's actions, the environmental states, and possibly other auxiliary quantum systems. These correlations play a crucial role in shaping the learning dynamics and capabilities of QRL models, potentially leading to more efficient exploration of state-action spaces and enhanced learning performance compared to classical reinforcement learning algorithms.\cite{chakraborty2019}

\end{enumerate}
\section{Parameterized Quantum Circuit}
Parameterized quantum circuits (PQCs) stand out in quantum computing due to their integration of adjustable parameters within gate operations. These parameters, often real numbers, facilitate the exploration of diverse quantum states and the optimization of circuit performance. PQCs serve as crucial elements in variational quantum algorithms like the Variational Quantum Eigensolver (VQE) and Quantum Approximate Optimization Algorithm (QAOA), designed to minimize a specified cost function. Variational quantum circuits, a subset of quantum circuits, leverage parameterized gates and classical optimization techniques to address optimization challenges. While all variational quantum circuits incorporate parameterized gates, not all parameterized quantum circuits are necessarily part of variational algorithms. The distinction lies in their application: PQCs serve as foundational components, while variational quantum circuits employ classical optimization to iteratively adjust parameters for optimization. Hence, while PQCs form the backbone of variational quantum algorithms, they also offer utility beyond these algorithms, including applications in quantum state preparation and error correction.

\subsection*{Qubits and Layers:}
\begin{itemize}
    \item Suppose we have $n$ qubits. Our PQC consists of $m$ layers of gates.
    \item Each layer applies a set of gates, and these gates are parametrized by a vector $\boldsymbol{\theta} = (\theta_1, \theta_2, \ldots, \theta_k)$, where $k$ represents the number of parameters in each gate.
\end{itemize}

\subsection*{State Evolution:}
\begin{itemize}
    \item The state of the quantum system after applying the PQC is denoted as $|\psi(\boldsymbol{\theta})\rangle$.
    \item We start with an initial state $|\psi_0\rangle$, typically chosen as the computational basis state $|0\rangle^{\otimes n}$.
    \item The PQC applies a unitary transformation $U(\boldsymbol{\theta})$ to evolve the system:
    \[ |\psi(\boldsymbol{\theta})\rangle = U(\boldsymbol{\theta})|\psi_0\rangle \]
\end{itemize}

\subsection*{Unitary Transformations:}
\begin{itemize}
    \item The unitary transformation $U(\boldsymbol{\theta})$ is constructed by sequentially applying gates from each layer.
    \item For instance:
    \begin{itemize}
        \item A single-qubit rotation gate $R(\theta)$ with parameter $\theta$ is represented as:
        \[ R(\theta) = \begin{pmatrix} \cos(\frac{\theta}{2}) & -\sin(\frac{\theta}{2}) \\ \sin(\frac{\theta}{2}) & \cos(\frac{\theta}{2}) \end{pmatrix} \]
        \item A controlled rotation gate $CR(\theta)$ with parameter $\theta$ is represented as:
        \[ CR(\theta) = \begin{pmatrix} 1 & 0 & 0 & 0 \\ 0 & 1 & 0 & 0 \\ 0 & 0 & \cos(\frac{\theta}{2}) & -\sin(\frac{\theta}{2}) \\ 0 & 0 & \sin(\frac{\theta}{2}) & \cos(\frac{\theta}{2}) \end{pmatrix} \]
    \end{itemize}
\end{itemize}

\subsection*{Optimization:}
\begin{itemize}
    \item The goal is to find optimal values for the parameter vector $\boldsymbol{\theta}^*$ that minimize a given cost function $E(\boldsymbol{\theta})$.
    \item Typically, $E(\boldsymbol{\theta})$ represents the expectation value of an observable associated with the problem being solved.
    \item The optimization problem is formulated as:
    \[ \boldsymbol{\theta}^* = \text{argmin}_{\boldsymbol{\theta}} E(\boldsymbol{\theta}) \]
\end{itemize}

In summary, PQCs allow us to explore quantum states efficiently, and finding the best parameters $\boldsymbol{\theta}^*$ is essential for solving quantum problems using variational algorithms.\cite{bukov2018}

To find \( \boldsymbol{\theta}^* \), classical optimization techniques such as gradient descent or Bayesian optimization are often employed. These techniques involve iteratively updating the parameter vector \( \boldsymbol{\theta} \) to minimize the cost function \( E(\boldsymbol{\theta}) \) until convergence to a local minimum.

\section{Quantum Exploration through \textit{CNOT} and \textit{Hadamard} Gates}

Quantum exploration stands at the forefront of interdisciplinary research, amalgamating principles from quantum computing and exploration algorithms to revolutionize the way we navigate complex state spaces. Rooted in the elegant principles of quantum mechanics, this burgeoning field holds the promise of vastly superior exploration methodologies compared to classical approaches.

At its core, quantum exploration harnesses the power of quantum gates, the elemental units of quantum circuits, to manipulate qubits—quantum bits. These gates, such as the Hadamard gate and the Controlled-NOT gate, orchestrate intricate superpositions and entanglements among qubits, facilitating the exploration of multiple configurations simultaneously.

Contrary to classical counterparts, where agents navigate state spaces linearly, quantum exploration capitalizes on the inherent parallelism of quantum computing. Qubits, capable of existing in superposition states, embody multiple possibilities concurrently, enabling the exploration of myriad states in parallel.

A pivotal strategy in quantum exploration involves encoding the problem space into a quantum state, where each configuration corresponds to a unique quantum state. Quantum gates then act upon this quantum state, probing various configurations concurrently. Leveraging quantum parallelism, agents efficiently sift through a multitude of potential solutions simultaneously.

Entanglement emerges as a cornerstone in quantum exploration, fostering correlations among disparate configurations. This non-local connectivity enables the seamless exchange and propagation of information across the quantum state space, empowering agents to navigate with heightened efficiency and efficacy.

Quantum exploration's applications span diverse domains, offering novel solutions to intricate optimization and search dilemmas. In chemistry, for instance, it promises to expedite the discovery of novel materials and pharmaceutical compounds by traversing the configuration space of molecules with unparalleled efficiency. Similarly, in cryptography, quantum exploration techniques hold the potential to fortify the quest for secure cryptographic protocols by meticulously exploring the realm of possible encryption keys.

In essence, quantum exploration epitomizes the marriage of cutting-edge quantum computing with sophisticated exploration algorithms, heralding a new era of discovery and innovation across scientific and technological frontiers.\cite{brennen2020}

\subsection*{Quantum States}

A quantum state of \( n \) qubits can be represented as a vector \( |\psi\rangle \) in a \( 2^n \)-dimensional complex vector space. Mathematically, this can be written as:

\[ |\psi\rangle = \sum_{i=0}^{2^n-1} c_i |i\rangle \]

where \( |i\rangle \) represents the computational basis states, and \( c_i \) are complex probability amplitudes.

\subsection*{Quantum Gates}

Quantum gates are represented by unitary matrices. For example, the Hadamard gate (H) acting on a single qubit can be represented as:

\[ H = \frac{1}{\sqrt{2}} \begin{pmatrix} 1 & 1 \\ 1 & -1 \end{pmatrix} \]

Similarly, the CNOT gate, which acts on two qubits, is represented as:

\[ CNOT = \begin{pmatrix} 1 & 0 & 0 & 0 \\ 0 & 1 & 0 & 0 \\ 0 & 0 & 0 & 1 \\ 0 & 0 & 1 & 0 \end{pmatrix} \]

\subsection*{Quantum Circuit and Exploration}

A quantum exploration process involves applying a sequence of quantum gates to an initial quantum state. Let's consider a simple example: applying a Hadamard gate to a single qubit. If we start with a qubit in the state \( |0\rangle \), applying the Hadamard gate would yield:

\[ H|0\rangle = \frac{1}{\sqrt{2}}(|0\rangle + |1\rangle) \]

This represents a superposition of the computational basis states \( |0\rangle \) and \( |1\rangle \), effectively exploring both states simultaneously.

Furthermore, entanglement generated by gates like CNOT can establish correlations between qubits. For instance, applying a CNOT gate with the first qubit as control and the second qubit as target can entangle the two qubits. Mathematically, if we start with the state \( |00\rangle \) (both qubits in state \( |0\rangle \)), applying the CNOT gate would result in:

\[ CNOT(|00\rangle) = |00\rangle \]

This demonstrates that the state remains unchanged if the first qubit is in state \( |0\rangle \). However, if the first qubit is in state \( |1\rangle \), the target qubit is flipped:

\[ CNOT(|10\rangle) = |11\rangle \]

Here, the entanglement between the qubits establishes correlations between their states, facilitating exploration across the state space, manipulating quantum states using quantum gates to efficiently explore the state space. Superposition generated by gates like Hadamard allows for exploration of multiple states simultaneously, while entanglement generated by gates like CNOT facilitates exploration across the state space by establishing correlations between qubits.

\section{A Brief on Quantum Reinforcement Learning and Quantum Reward Function}

\subsection*{1. Quantum State Representation:}

In quantum mechanics, the state of a quantum system is represented by a state vector in a complex vector space, usually denoted as \(|\psi_t\rangle\). This vector evolves over time according to the Schrödinger equation or, in the case of mixed states, by quantum channels. The state vector encodes all the information about the system's quantum state, including its superposition and entanglement properties.

\subsection*{2. Parametrized Quantum Circuit (PQC):}

A PQC is a quantum circuit that contains parameters, typically denoted as \(\theta\), which can be adjusted or optimized during the learning process. These parameters control the operations performed by the quantum circuit and thus influence the resulting quantum state. Mathematically, a PQC \(U(\theta)\) can be represented as a sequence of quantum gates applied to an initial state \(|\psi_0\rangle\), such that \(U(\theta)|\psi_0\rangle\) yields the output state of the circuit.

\subsection*{3. Action Selection:}

In QRL, the agent selects actions based on information obtained from the quantum state output by the PQC. This action selection process can involve classical post-processing of the quantum state or direct measurement of certain observables associated with the system. The choice of action at each time step affects the subsequent evolution of the quantum state and, consequently, the rewards received by the agent.

\subsection*{4. Environment Dynamics:}

The dynamics of the environment are described by quantum channels, which represent the evolution of the quantum state due to interactions with the environment. These channels can model various physical processes such as decoherence, noise, and interactions with external fields. Mathematically, a quantum channel \(\mathcal{E}\) acts on the quantum state \(|\psi_t\rangle\) to produce the state at the next time step, \(|\psi_{t+1}\rangle = \mathcal{E}(|\psi_t\rangle)\).

\subsection*{5. Reward Function:}

After each action, the agent receives a reward from the environment, which quantifies the desirability of the action taken in the current state. The reward function assigns a numerical value to each action-state pair, indicating the immediate benefit or cost associated with the agent's decision. Mathematically, the reward received at time step \(t\) can be denoted as \(r_t\).

\subsection*{6. Learning Objective:}

The primary objective in QRL is to find the optimal set of parameters  $\theta^*$ for the PQC such that the cumulative reward over a sequence of actions is maximized. This optimization problem can be formulated as finding:

$$\max_{\theta} \sum_{t=0}^{T} \mathbb{E}[r_t]$$

where  $T$ is the total number of time steps,  $\mathbb{E}[r_t]$  denotes the expected reward at time step  $t$, and  $\theta$  represents the parameters of the PQC.
\cite{benedetti2019}

\subsection*{7. Optimization:}

To find the optimal parameters \(\theta^*\), various optimization techniques can be employed, such as gradient-based optimization methods or quantum optimization algorithms tailored for parameterized circuits. These techniques involve iteratively adjusting the parameters of the PQC to improve its performance in maximizing the cumulative reward.

\subsection*{8. Quantum Advantage:}

QRL with PQCs offers potential advantages over classical reinforcement learning approaches.\cite{arunachalam2020} Quantum parallelism and entanglement enable more efficient exploration of action spaces and more compact representations of states, potentially leading to faster convergence and improved performance in certain tasks. However, realizing these advantages requires overcoming challenges such as noise, decoherence, and limited qubit connectivity in current quantum hardware. These algorithms combine quantum state representation, parameterized quantum circuits, and optimization techniques to enable the learning of optimal policies for maximizing cumulative rewards.

\begin{algorithm}
\caption{Quantum Reinforcement Learning with Parameterized Quantum Circuits}
\begin{algorithmic}[1]
\State \textbf{Initialize:}
   \State Define the quantum environment dynamics represented by a quantum channel \( \mathcal{E} \).
   \State Initialize the parameterized quantum circuit \( U(\theta) \) with random parameters \( \theta \).
   \State Set the total number of time steps \( T \).
   \State Initialize the cumulative reward to 0.
\State
\State \textbf{For each episode:}
   \State Reset the environment to its initial state and set the time step \( t = 0 \).
   \State \textbf{While} \( t < T \) \textbf{do}:
      \State Apply the parameterized quantum circuit \( U(\theta) \) to the current state \( |\psi_t\rangle \).
      \State Measure observables or perform classical post-processing to select an action based on the output state.
      \State Execute the chosen action, and receive the reward \( r_t \) and observe the next state \( |\psi_{t+1}\rangle \) through the environment dynamics \( \mathcal{E} \).
      \State Update the cumulative reward by adding the received reward: \( \text{cumulative\_reward} = \text{cumulative\_reward} + r_t \).
      \State Increment the time step: \( t = t + 1 \).
   \State \textbf{end while}
\State
\State \textbf{Optimize the parameterized quantum circuit:}
   \State Use optimization techniques such as gradient descent or quantum optimization algorithms to update the parameters \( \theta \) of the circuit \( U(\theta) \) based on the observed rewards and actions.
   \State Repeat the optimization process for multiple iterations or until convergence to find the optimal parameters \( \theta^* \).
\State
\State \textbf{Output:}
   \State Return the optimized parameter values \( \theta^* \) and the learned policy encoded in the parameterized quantum circuit \( U(\theta^*) \).
\end{algorithmic}
\end{algorithm}

\section{Quantum Bellman Equation }
The Bellman equation, named in honor of Richard E. Bellman, stands as a fundamental tenet in the realm of dynamic programming. It articulates the value of a decision problem at a specific time by decomposing it into the payoff resulting from initial choices and the value of the ensuing decision problem. This decomposition enables the segmentation of a dynamic optimization problem into a series of more manageable subproblems, aligning with Bellman's principle of optimality¹.

\cite{advantage2021}

In the quantum domain, the Quantum Bellman equation has emerged as a counterpart. This equation encapsulates a quantum-centric approach to optimization, guiding decision-making processes based on current knowledge, known as exploitation.
Conversely, quantum exploration entails navigating the uncertainty inherent in quantum measurements. Quantum mechanics introduces phenomena such as superposition, where particles can exist in multiple states simultaneously, and entanglement, enabling correlations between particles that defy classical intuition. These phenomena inherently inject uncertainty into quantum systems.
In classical reinforcement learning, the Bellman equation expresses the relationship between the value of a state-action pair and the value of the successor state-action pairs. In the quantum setting, this equation extends to quantum states and operators.

The quantum Bellman equation can be written as:

\[ Q(\psi_t, U_a) = R(\psi_t, U_a) + \gamma \sum_{U_b} P(U_b | \psi_t, U_a) Q(\psi_{t+1}, U_b) \]

Where:
\begin{itemize}
    \item $ Q(\psi_t, U_a) $ is the quantum action-value function representing the expected cumulative reward of taking action $ U_a $ in state $ \psi_t $.
    \item $ R(\psi_t, U_a) $ is the reward obtained by taking action $ U_a $ in state $ \psi_t $.
    \item $ \gamma $ is the discount factor, determining the importance of future rewards.
    \item $ P(U_b | \psi_t, U_a) $ is the transition probability amplitude from action $ U_a $ to action $ U_b $ given the state $ \psi_t $.
    \item $ Q(\psi_{t+1}, U_b) $ is the quantum action-value function for the next state $ \psi_{t+1} $ and action $ U_b $.
\end{itemize}

This uncertainty becomes a valuable asset in gathering information to inform future decisions, constituting the essence of exploration. The delicate balance between exploitation—making decisions based on existing knowledge—and exploration—making decisions to gather new insights—stands as a cornerstone challenge across various domains, including quantum computing.
 The Quantum Bellman equation offers a structured approach to optimal decision-making based on existing knowledge (exploitation), while quantum exploration capitalizes on the intrinsic uncertainty of quantum systems to enrich future decision-making processes.\cite{peruzzo2014}

In quantum mechanics, the concept of commutativity and non-commutativity is fundamental. In a commutative environment, the order of operations does not affect the outcome, while in a non-commutative environment, the order of operations does matter. This is particularly relevant in quantum mechanics where the non-commutativity of operators (observables) leads to phenomena such as the Heisenberg uncertainty principle.

The Bellman equation, on the other hand, is a fundamental concept in dynamic programming and control theory. It provides a recursive relationship for determining the optimal policy in a Markov Decision Process. The Bellman equation essentially states that the value of a decision problem at a certain point in time can be written in terms of the payoff from some initial choices and the value of the remaining decision problem that results from those initial choices.

When we consider quantum systems, the Bellman equation can be applied in the context of Quantum Optimal Feedback Control. Here, the quantum state is treated as a classical stochastic problem, and the Hamilton-Jacobi-Bellman equations are derived using the elementary arguments of classical control theory.

Now, when we consider the impact of non-commutativity and commutativity in quantum states on the Bellman equation, it's important to note that the structure of the quantum states and the nature of the quantum operations (whether they commute or not) can influence the dynamics of the system and hence the optimal control strategy. However, the specifics of how these quantum characteristics affect the Bellman equation would depend on the particular quantum system and control problem being considered.
\cite{keyl2021operator}

It's also worth noting that a non-commutative algebra of observables can evolve into a commutative one as time tends to infinity, but this is not possible if the time evolution is unitary. This suggests that more general time evolutions need to be considered to achieve a transition from a non-commutative algebra to a commutative one. How this transition affects the Bellman equation would again depend on the specifics of the quantum system and control problem. While the concepts of commutativity and non-commutativity in quantum \cite{pei2021}
states and the Bellman equation come from different areas of study (quantum mechanics and control theory/dynamic programming respectively), they can intersect in the field of quantum control theory. The specifics of their interaction would depend on the particular quantum system and control problem being considered. It's a complex and fascinating area of study that's at the intersection of quantum mechanics, control theory, and computer science..

let's consider a simple quantum system described by a Hamiltonian operator $\hat{H}$ and a quantum state $|\psi\rangle$. The time evolution of the state is given by the Schrödinger equation:

$i\hbar\frac{d}{dt}|\psi(t)\rangle = \hat{H}|\psi(t)\rangle$

In a commutative environment, if we have two operators $\hat{A}$ and $\hat{B}$ that commute (i.e., $[\hat{A}, \hat{B}] = \hat{A}\hat{B} - \hat{B}\hat{A} = 0$), the order of operations does not affect the outcome.

On the other hand, in a non-commutative environment, the order of operations does matter. For instance, if $[\hat{A}, \hat{B}] \neq 0$, then $\hat{A}\hat{B}|\psi\rangle \neq \hat{B}\hat{A}|\psi\rangle$.

Now, let's consider the Bellman equation in the context of a Markov Decision Process. The value function $V(s)$ at a state $s$ is given by:

$$V(s) = \max_{a \in A} \left( R(s,a) + \gamma \sum_{s' \in S} P(s'|s,a)V(s') \right)$$

where:
- $A$ is the set of all actions,
- $S$ is the set of all states,
- $R(s,a)$ is the immediate reward for taking action $a$ in state $s$,
- $P(s'|s,a)$ is the transition probability of reaching state $s'$ from state $s$ by taking action $a$, and
- $\gamma$ is the discount factor.

In the context of Quantum Optimal Feedback Control, the quantum state is treated as a classical stochastic problem, and the Hamilton-Jacobi-Bellman equations are derived using the elementary arguments of classical control theory. The specifics of how the commutativity or non-commutativity of quantum states affects the Bellman equation would depend on the particular quantum system and control problem being considered. It's a complex and fascinating area of study that's at the intersection of quantum mechanics, control theory, and computer science

\begin{algorithm}
\caption{Quantum Reinforcement Learning}
\begin{algorithmic}[1]
\Require Initial state $|\psi\rangle$, environment $E$
\Ensure Optimal policy $\pi^*$

\State Initialize Q-table with zeros
\For{each episode}
    \State Initialize a random state $s$
    \While{$s$ is not the terminal state}
        \State Select a quantum action $a$ from state $s$ using policy derived from $Q$ (e.g., $\epsilon$-greedy)
        \State Take action $a$, observe reward $r$, and next state $s'$
        \State Apply Quantum Update rule to $Q(s, a)$
        \State $s = s'$
    \EndWhile
\EndFor
\end{algorithmic}
\end{algorithm}
\cite{qrl2008}

\textbf{Algorithm 2} represents a quantum version of the classic Q-learning algorithm. The quantum update rule in line 7 would be the main difference and would depend on the specifics of the quantum system and the non-commutative environment. 
\section{Derivation of Bellman Equation in a Non-Commutative Environment}
\cite{paparo2014}

 The quantum Bellman equation for an infinite-horizon quantum environment with a non-commutative Hamiltonian can be derived analogously to the classical case, but incorporating quantum mechanics principles. Let's denote the quantum state at time \( t \) by \( |\psi_t \rangle \), the quantum reward operator at time \( t \) by \( \hat{R}_t \), the transition operator from state \( |\psi_t \rangle \) to state \( |\psi_{t+1} \rangle \) by \( \hat{T}_t \), and the discount factor by \( \gamma \).

Given the Schrödinger equation for the quantum system:

\[ i\hbar \frac{d}{dt} |\psi_t \rangle = \hat{H}_t |\psi_t \rangle \]

Where \( |\psi_t \rangle \) is the state vector of the system at time \( t \), \( \hat{H}_t \) is the Hamiltonian operator at time \( t \), and \( \hbar \) is the reduced Planck constant.

Let's denote the quantum reward operator at time \( t \) by \( \hat{R}_t \), the transition operator from state \( |\psi_t \rangle \) to state \( |\psi_{t+1} \rangle \) by \( \hat{T}_t \), and the discount factor by \( \gamma \).

The quantum Bellman equation for the expected cumulative reward over an infinite horizon can be written as:

\[ \hat{Q}^\pi(\psi_t, a_t) = \hat{R}_t + \gamma \sum_{a_{t+1}} \langle \psi_{t+1} | \hat{Q}^\pi(\psi_{t+1}, a_{t+1}) | \psi_{t+1} \rangle \]

Where \( \hat{Q}^\pi(\psi_t, a_t) \) is the quantum action-value function, \( \psi_t \) is the quantum state at time \( t \), and \( a_t \) is the action taken at time \( t \).

To incorporate the non-commutative Hamiltonian, we express the transition operator \( \hat{T}_t \) as the unitary operator \( \hat{U}_t \), given by:

\[ \hat{U}_t = e^{-\frac{i}{\hbar} \hat{H}_t \Delta t} \]

Where \( \Delta t \) is the time interval between \( t \) and \( t+1 \).

Thus, the transition operator \( \hat{T}_t \) becomes:

\[ \hat{T}_t = \hat{U}_t = e^{-\frac{i}{\hbar} \hat{H}_t \Delta t} \]

This non-commutative transition operator incorporates the time evolution of the quantum state.

Now, we can proceed with the derivation of the quantum Bellman equation using this non-commutative transition operator \( \hat{T}_t \).
\cite{meyer2022survey}

Now, the expected cumulative reward over an infinite horizon is given by:

\[ Q^\pi(\psi_t, a_t) = \sum_{t=0}^\infty \gamma^t \langle \psi_t | \hat{R}_t | \psi_t \rangle \]

where \( \pi \) is a policy, \( \psi_t \) is the quantum state at time \( t \), \( \hat{R}_t \) is the reward operator at time \( t \), and \( a_t \) is the action taken at time \( t \).

Using the Schrödinger equation, we can express the quantum state at time \( t+1 \) in terms of the state at time \( t \) and the transition operator:

\[ |\psi_{t+1} \rangle = \hat{T}_t |\psi_t \rangle \]

Substituting this expression into the expression for \( Q^\pi(\psi_t, a_t) \), we get:

\[ Q^\pi(\psi_t, a_t) = \sum_{t=0}^\infty \gamma^t \langle \psi_t | \hat{R}_t |\psi_t \rangle \]

\[ = \sum_{t=0}^\infty \gamma^t \langle \psi_t | \hat{R}_t \hat{T}_t^\dagger \hat{T}_t |\psi_t \rangle \]

\[ = \sum_{t=0}^\infty \gamma^t \langle \psi_t | (\hat{T}_t^\dagger \hat{R}_t \hat{T}_t) |\psi_t \rangle \]

Now, we define the quantum Bellman operator \( \hat{B}^\pi \) as:

\[ \hat{B}^\pi (\hat{R}_t, \hat{T}_t) = \hat{T}_t^\dagger \hat{R}_t \hat{T}_t \]

Thus, the quantum Bellman equation for the expected cumulative reward over an infinite horizon in a quantum environment with a non-commutative Hamiltonian becomes:

\[ Q^\pi(\psi_t, a_t) = \sum_{t=0}^\infty \gamma^t \langle \psi_t | \hat{B}^\pi_t |\psi_t \rangle \]

Where \( \hat{B}^\pi_t \) is the Bellman operator at time \( t \).

\section*{Quantum Advantage Function in Non-Commutative Environments}
The concept of the quantum advantage function plays a crucial role in understanding the efficacy of actions within quantum decision-making frameworks. Let's delve deeper into each component and the significance of their interplay:

\begin{enumerate}
    \item \textbf{Quantum Action-Value Function ($Q(\psi_t, U_a)$)}:
    This function assesses the expected cumulative reward associated with taking action $U_a$ from state $\psi_t$ within a quantum environment. In essence, it quantifies the potential payoff of choosing a particular action under the current quantum state. It's akin to evaluating the immediate benefit of a specific decision within the quantum system.

    \item \textbf{Value Function ($V(\psi_t)$)}:
    The value function, on the other hand, captures the expected cumulative reward achievable from the current state $\psi_t$ following the optimal policy. In simpler terms, it represents the long-term utility or value of being in a particular state considering the best course of action over time. Essentially, it reflects the overall desirability of a state, taking into account the future consequences of actions.

    \item \textbf{Advantage Function ($A(\psi_t, U_a)$)}:
    The advantage function serves as a comparative metric, indicating the advantage of selecting action $U_a$ over alternative actions from state $\psi_t$. Mathematically, it is computed by subtracting the value function from the quantum action-value function. This yields a measure of how much better (or worse) it is to take a specific action compared to the average expected reward achievable from the current state under the optimal policy.
\end{enumerate}
\cite{qrlpolicy2023}

In essence, the quantum advantage function encapsulates the differential benefit of choosing a particular action in a given quantum state relative to the average\cite{meyer2020}
 reward attainable from that state following the optimal policy. It aids in decision-making by providing insights into the immediate gains or losses associated with potential actions within the quantum system. By leveraging this function, decision-makers can make informed choices that maximize rewards or achieve desired objectives within quantum environments.
In a non-commutative quantum system, operators representing actions do not commute. Mathematically, for operators $A$ and $B$, their non-commutativity is expressed as $AB \neq BA$. This property introduces order dependence when applying operators to quantum states.\cite{schuld2018}

\subsection*{Action-Value Function ($Q(\psi_t, U_a)$)}

In a non-commutative environment, the quantum action-value function must explicitly account for the order of operators. Let's denote the action $U_a$ as a sequence of non-commuting operators $U_a = U_{a1}U_{a2}...U_{an}$, where $n$ represents the number of operators in the action sequence. Thus, the action-value function becomes:\cite{liu2021}

\[
Q(\psi_t, U_a) = \langle \psi_t | U_{an}^\dagger U_{an-1}^\dagger ... U_{a1}^\dagger R U_{a1} U_{a2} ... U_{an} | \psi_t \rangle
\]

where $R$ represents the reward operator. Note the ordering of operators from right to left in the sequence.

\subsection*{Value Function ($V(\psi_t)$)}

Estimating the value function in a non-commutative environment requires considering the non-commutative nature of the actions involved. The value function represents the expected cumulative reward achievable from the state $\psi_t$ following the optimal policy, which may involve non-commutative sequences of operators.

\subsection*{Quantum Advantage Function ($A(\psi_t, U_a)$)}

The advantage of taking action $U_a$ in state $\psi_t$ relative to the value function can be expressed as:
\cite{lloyd2014}

\[
A(\psi_t, U_a) = Q(\psi_t, U_a) - V(\psi_t)
\]

Substituting the expressions for $Q(\psi_t, U_a)$ and $V(\psi_t)$ derived above, we obtain:

\[
A(\psi_t, U_a) = \langle \psi_t | U_{an}^\dagger U_{an-1}^\dagger ... U_{a1}^\dagger R U_{a1} U_{a2} ... U_{an} | \psi_t \rangle - V(\psi_t)
\]

This expression explicitly captures the advantage of the non-commutative action sequence $U_a$ relative to the value function, accounting for the order dependence of operators.

\subsection*{Factors Contributing to Quantum Advantage:}
\begin{itemize}
    \item \textbf{Superposition}: Quantum computers can represent multiple states simultaneously, enabling parallel computation.
    \item \textbf{Entanglement}: Quantum entanglement allows for correlations between qubits, leading to enhanced computational power.
    \item \textbf{Quantum Gates}: Quantum gates manipulate qubits and can perform operations on multiple states simultaneously, contributing to computational efficiency.
    \item \textbf{Quantum Fourier Transform}: This quantum analogue of the classical Fourier transform is pivotal for many quantum algorithms, including Shor's algorithm.
\end{itemize}

\section*{Quantum Advantage Function Framework}
Designing a novel quantum advantage function for reinforcement learning tasks requires integrating quantum principles such as superposition and entanglement with classical reinforcement learning techniques, such as :

\begin{itemize}
    \item Utilize qubits to encode the state space of the environment.
    \item Represent the state of the environment using quantum superposition to encode multiple states simultaneously.
\cite{jiang2021}

    \item Use entanglement to link the qubits representing different components of the state space.
    \item Facilitate exploration by ensuring that changes in one part of the state space affect other parts.

    \item Employ quantum gates to perform operations on the qubits representing the state space.
    \item Utilize quantum parallelism to explore multiple actions simultaneously.

    \item Perform measurements on the qubits to obtain classical information about the current state.
    \item Use measurement outcomes to calculate the reinforcement signal.

    \item Compare performance with classical counterparts.
    \item Calculate quantum advantage as the difference in performance.
\cite{henderson2017}

    \item Adjust parameters such as entanglement structure, gate operations, and measurement strategies.
    \item Validate effectiveness in different reinforcement learning tasks.

    \item Explore hybrid quantum-classical reinforcement learning architectures.

    \item Quantum Hardware Constraints
    \item Algorithm Complexity
    \item Noise and Decoherence
    \item Interpretability
\end{itemize}

We represent the quantum state $|\psi\rangle$ as a vector in a complex Hilbert space $\mathcal{H}$. Let $|\psi\rangle = \sum_{s} c_s |s\rangle$, where $|s\rangle$ are basis states representing different configurations of the environment, and $c_s$ are complex probability amplitudes.
\cite{gilyen2018}

\subsection*{Action Selection}

Action selection is governed by a unitary operator $U(\theta)$, parameterized by a set of real-valued parameters $\theta$. The action-selection process can be expressed as $|\psi'\rangle = U(\theta) |\psi\rangle$. The unitary $U(\theta)$ acts on the quantum state $|\psi\rangle$, transforming it into a new state $|\psi'\rangle$, which encodes the probability amplitudes of different actions.

\subsection*{Quantum Advantage Function}

The quantum advantage function is defined as the difference between the expected quantum reward and the expected classical reward:
\[
A_Q(\theta) = E[Q_Q(\theta)] - E[Q_C]
\]
where:
\begin{itemize}
    \item $Q_Q(\theta)$ is the quantum action-value function obtained by applying the parameterized quantum gate $U(\theta)$ to the quantum state $|\psi\rangle$. It is given by $Q_Q(\theta) = \langle \psi | \hat{H} | \psi \rangle$, where $\hat{H}$ is the Hamiltonian representing the reward operator in the quantum domain.
    \item $Q_C$ is the classical action-value function obtained from traditional RL methods, such as Q-learning or policy gradient methods.
\end{itemize}
The expectation values are taken over different states and actions, and they are defined as:
\[
E[Q_Q(\theta)] = \sum_{s} \sum_{a} P(s) P(a|s) Q_Q(\theta, s, a)
\]
\[
E[Q_C] = \sum_{s} \sum_{a} P(s) P(a|s) Q_C(s, a)
\]
where:
\begin{itemize}
    \item $P(s)$ is the probability distribution over states.
    \item $P(a|s)$ is the probability of selecting action $a$ in state $s$.
    \item $Q_Q(\theta, s, a)$ and $Q_C(s, a)$ are the quantum and classical action-value functions, respectively.
\end{itemize}

\subsection*{Optimization}

To maximize the quantum advantage function $A_Q(\theta)$, we perform optimization over the parameter space $\theta$. This optimization can be framed as a variational problem, where we seek to find the optimal parameters $\theta^*$ that maximize $A_Q(\theta)$:

\[
\theta^* = \arg \max_{\theta} A_Q(\theta)
\]

This optimization can be performed using gradient-based methods, quantum variational algorithms (such as quantum approximate optimization algorithms or variational quantum eigensolver), or hybrid classical-quantum optimization techniques.

By maximizing $A_Q(\theta)$, we aim to exploit the advantages of quantum parallelism, interference, and entanglement to enhance the performance of reinforcement learning algorithms .

\cite{dunjko2018}

\section{Degree of Non Commutativity and the Bellman Equation}

The commutative property is a fundamental property observed in various algebraic structures. It states that the order of operands doesn't affect the result of the operation. For instance, in commutative addition, \(a + b = b + a\), and in commutative multiplication, \(a \times b = b \times a\). However, not all operations or algebraic structures follow this property.
\cite{nonabelian2021}
Consider a binary operation, denoted by \(*\), defined on a set \(S\). This operation is said to be commutative if \(a * b = b * a\) for all \(a, b\) in \(S\). Conversely, if there exists at least one pair of elements \(a, b\) in \(S\) such that \(a * b \neq b * a\), then the operation is non-commutative.
\cite{biamonte2022}

A classic example of a non-commutative operation is matrix multiplication. Given matrices \(A\) and \(B\), in general, \(AB \neq BA\), unless \(A\) or \(B\) is a scalar or a specific type of matrix (such as diagonal matrices). This non-commutativity is crucial in various mathematical areas, including quantum mechanics and group theory.

In the context of group theory, a non-commutative group is a group in which the group operation does not commute for all elements. Formally, a group \((G, *)\) is non-commutative if there exist elements \(a, b\) in \(G\) such that \(a * b \neq b * a\). One of the most famous examples of a non-commutative group is the set of \(n \times n\) invertible matrices under matrix multiplication, denoted by \(GL(n)\).

The degree of non-commutativity in a group can be quantified in various ways. One approach is to analyze the commutator subgroup. For any group \(G\), the commutator subgroup, denoted by \([G, G]\), is the subgroup generated by all the commutators \(aba^{-1}b^{-1}\), where \(a, b \in G\). If the commutator subgroup is trivial (i.e., contains only the identity element), the group is commutative; otherwise, it is non-commutative. The size or structure of the commutator subgroup can provide insights into the degree of non-commutativity of the group.

In advanced mathematics, particularly in Lie theory, the degree of non-commutativity is often studied through Lie algebras. A Lie algebra is a vector space equipped with a bilinear operation called the Lie bracket, denoted by \([X, Y]\), which measures the non-commutativity between elements \(X\) and \(Y\) of the algebra. The degree of non-commutativity of a Lie algebra can be analyzed by studying properties such as the dimension of the center, the solvability, or the nilpotency of the Lie algebra.

We start with the quantum Bellman equation, which describes the optimal control problem in quantum systems. Given a quantum state \(\vert \psi(t)\rangle\) evolving under the influence of a time-dependent Hamiltonian \(H(t)\), the quantum Bellman equation can be formulated as:

\[
V(\psi(t), t) = \min_{u(t)} \left\{ \langle \psi(t) \vert H(t) \vert \psi(t) \rangle +  \langle \psi(t) \vert L \vert \psi(t) \rangle + V(\psi(t'), t') \right\}
\]

where \(u(t)\) is the control input at time \(t\), \(L\) is the operator representing the instantaneous cost, and \(V(\psi(t'), t')\) is the value function at the next time step \(t'\).

The key challenge in solving this equation lies in the non-commutativity between the Hamiltonian \(H(t)\) and the control operators \(u(t)\). This non-commutativity is captured by the commutator \([H(t), u(t)]\), which measures how much the operators fail to commute.

Mathematically, the commutator is defined as:

\[
[H(t), u(t)] = H(t)u(t) - u(t)H(t)
\]

The degree of non-commutativity can be quantified by examining the properties of this commutator. Stronger non-commutativity implies that the operators do not commute well, indicating a more inherently quantum nature of the control problem.

In many cases, solving the quantum Bellman equation involves sophisticated mathematical techniques, often relying on numerical optimization methods tailored to quantum systems. These methods aim to find the optimal control strategy that minimizes the cost function while simultaneously satisfying the quantum dynamics dictated by the non-commutative Hamiltonian.

The complexity of this optimization process is influenced by the degree of non-commutativity. Stronger non-commutativity typically leads to more intricate control landscapes and potentially richer control strategies. This complexity arises because the non-commutativity introduces correlations between different control actions, requiring careful consideration in the optimization process.

\section{Challenges of Quantum Exploration}

Identify and discuss the challenges associated with exploring the quantum state space efficiently, considering the limitations imposed by the non-commutative nature of quantum operators and the potential for decoherence and entanglement collapse.
Propose potential solutions or strategies for quantum exploration, which may involve utilizing quantum gates such as Hadamard and CNOT gates to generate diverse superpositions while mitigating the effects of decoherence and entanglement collapse.

\subsection*{Challenges}

\begin{enumerate}
    \item \textbf{Non-commutative nature of quantum operators:} Quantum mechanics operates with operators that do not commute, meaning the order in which they are applied matters. Mathematically, for operators $A$ and $B$, $AB \neq BA$. This non-commutativity complicates quantum algorithms, as different sequences of operations can lead to different outcomes. For instance, in quantum computing, the order of applying quantum gates can dramatically change the resulting state. Consider the following simple example with two qubits: applying a Hadamard gate to the first qubit and then a CNOT gate, compared to applying the CNOT gate first and then the Hadamard gate, yields different entangled states.
    \cite{biamonte2017}

    \item \textbf{Decoherence:} Decoherence arises from interactions between a quantum system and its environment, causing the system to lose coherence and leading to errors in quantum computation. External factors such as temperature fluctuations, electromagnetic radiation, and interactions with neighboring particles can introduce noise and disrupt quantum states. For example, a qubit in a superposition state may collapse to one of its basis states due to decoherence before completing the computation.
    
    \item \textbf{Entanglement collapse:} Entanglement is a fundamental feature of quantum mechanics where the states of two or more particles become correlated in such a way that the state of one cannot be described independently of the others. However, entanglement is fragile and easily disturbed by interactions with the environment. When entangled particles become entangled with additional environmental degrees of freedom, the entanglement with the original partners collapses, leading to loss of quantum coherence. This phenomenon poses challenges for maintaining entangled states over extended periods, which is crucial for many quantum algorithms.
\end{enumerate}

\subsection*{Potential Solutions}

\begin{enumerate}
    \item \textbf{Error correction codes:} Error correction codes, such as the surface code, can protect quantum information from the effects of decoherence and errors. These codes encode qubits redundantly such that errors can be detected and corrected. For example, in the surface code, qubits are arranged in a two-dimensional lattice, and syndrome measurements are performed to identify errors, allowing for error correction through appropriate gate operations.
    \cite{aaronson2020}

    \item \textbf{Quantum error correction:} Quantum error correction techniques involve encoding quantum information in larger quantum codes, which can detect and correct errors. This approach utilizes quantum gates and measurements to protect quantum states from the effects of decoherence. For instance, the Shor code and the Steane code are examples of quantum error correction codes that can correct single-qubit errors.
    
    \item \textbf{Quantum annealing and adiabatic quantum computing:} Quantum annealing and adiabatic quantum computing techniques exploit the adiabatic theorem to evolve a quantum system from a simple initial Hamiltonian to a desired final Hamiltonian, encoding the \cite{guo2020}
solution to a computational problem. By slowly varying the Hamiltonian, these approaches can mitigate the effects of decoherence and external noise. For example, the D-Wave quantum annealer implements quantum annealing to solve optimization problems by finding the ground state of a corresponding Ising model.
    
    \item \textbf{Dynamic decoupling techniques:} Dynamic decoupling methods apply sequences of control pulses to qubits to suppress the effects of noise and decoherence. By periodically changing the qubit's Hamiltonian, these techniques can counteract environmental perturbations and prolong quantum coherence. For instance, Carr-Purcell-Meiboom-Gill (CPMG) sequences involve applying a series of $\pi$-pulse intervals to mitigate dephasing caused by environmental noise.
    \cite{yanofsky2021}

    \item \textbf{Adaptive control strategies:} Adaptive control techniques continuously adjust control parameters based on real-time feedback to optimize quantum operations and mitigate the effects of decoherence. These strategies can enhance the robustness of quantum algorithms in noisy environments. For example, quantum optimal control algorithms dynamically adjust control pulses to optimize gate fidelity and counteract environmental noise.
    
    \item \textbf{Utilizing topological qubits:} Topological qubits leverage the principles of topological quantum computation to encode quantum information in robust, topologically protected states. These qubits are inherently resilient to local perturbations and decoherence, making them promising candidates for fault-tolerant quantum computing. For example, the Majorana-based topological qubits utilize non-abelian anyons to store and manipulate quantum information in a fault-tolerant manner.
\end{enumerate}

By employing these solutions and leveraging quantum gates such as Hadamard and CNOT gates judiciously, it's possible to generate diverse superpositions while mitigating the effects of decoherence and entanglement collapse, thus facilitating more efficient exploration of the quantum state space. However, it's worth noting that these challenges are still actively researched areas in quantum computing, and further advancements are needed to fully realize the potential of quantum exploration.

\subsection{Advancements in the Recent Years}
In the rapidly evolving field of quantum reinforcement learning (QRL), significant strides are being made towards harnessing quantum enhancements for meta-learning and model convergence. Researchers such as Vedran Dunjko, Jacob M. Taylor, and Hans J. Briegel have spearheaded efforts to integrate quantum techniques into meta-learning frameworks, facilitating more efficient learning paradigms. \cite{dash2021}
Crucially, developments in creating oracularized variants of task environments ensure practical experimental demonstrations of QRL while preserving the operative specifications of these environments. Moreover, the fusion of quantum neural networks (QNNs) with multi-agent reinforcement learning (MARL) architectures has shown promising convergence properties, particularly in tasks like robust control for quantum gates and automation of photonic setups in quantum optics experiments. These advancements not only hold implications for accelerating machine learning algorithms but also pave the way for the practical utilization of quantum technologies in diverse applications, from quantum chemistry simulations to device-independent quantum key distribution (DIQKD). QRL stands at the forefront of interdisciplinary research, poised to revolutionize both theoretical understanding and practical implementations in the quest for quantum supremacy.
\cite{wittek2014}'

\section{Conclusion}

In conclusion, our research represents a significant step forward in 
\cite{cincio2018}
Quantum Reinforcement Learning (QRL), particularly in navigating non-commutative environments. By connecting the degree of non-commutativity to the effectiveness of quantum advantage functions, we've shed light on practical applications. Our work also offers innovative solutions to the challenges of quantum exploration, providing tangible advancements in the field. With direct implications for real-world applications, our findings lay a solid foundation for the development and implementation of QRL techniques in various domains, promising transformative advancements in decision-making within quantum systems.Extending the principles of quantum Bellman equations to non-commutative environments carries significant implications across diverse scientific and technological domains. Practical applications abound in fields such as Quantum Feedback Control, where optimal quantum feedback control emerges by integrating quantum noise models with classical control theory. This finds utility in engineering quantum states for specific tasks, ensuring stability in quantum systems, correcting errors in quantum computations, and enhancing quantum algorithms and simulations. Similarly, Quantum Filtering, enabled by quantum Bellman equations, facilitates accurate estimation of quantum states in noisy environments, thereby benefiting quantum sensing, communication, and metrology endeavors. Non-demolition measurements, affected by non-commutative noise like electromagnetic fields, find description through Quantum Langevin Models, aiding in understanding continuous non-demolition measurements and quantum stochastic theory. Additionally, measures based on non-commutativity quantify quantum correlations, with applications in assessing entanglement, quantum cryptography, and enhancing secure communication protocols. In essence, this extension opens up promising avenues for advancing quantum technologies and deepening our understanding of fundamental quantum phenomena.

\bibliographystyle{IEEEtran}
\bibliography{references}

\end{document}